\documentclass[aps,preprint,amsfonts,amssymb,groupedaddress]{revtex4}
\usepackage[english]{babel}
\usepackage{epsfig}
\usepackage{amsfonts, amssymb, amsmath}
\begin{document}

\title{Theory of $4e$ versus $2e$ supercurrent in  frustrated Josepshon-junction rhombi chain}

\author{Ivan V. Protopopov and Mikhail V. Feigel'man}

\affiliation{L.D.Landau Institute for Theoretical Physics,Kosygin str.2,  Moscow, 119334,
Russia}

\begin{abstract}
We consider  a chain of  Josepshon-junction rhombi (proposed originally in~\cite{Doucot}) in quantum
regime, and in the realistic case when charging effects are determined by junction
capacitances. In the maximally frustrated case when magnetic flux through each rhombi
$\Phi_r$ is equal to one half of superconductive flux quantum $\Phi_0$, Josepshon current
is due to correlated transport of {\em pairs of Cooper pairs}, i.e. charge is quantized
in units of $4e$. Sufficiently strong deviation $ \delta\Phi \equiv |\Phi_r-\Phi_0/2| > \delta\Phi^c$
from the maximally frustrated point brings the system back to usual $2e$-quantized supercurrent.
 We present detailed analysis of Josepshon current in the fluctuation-dominated
regime (sufficiently long chains) as function of the chain length, $E_J/E_C$ ratio and
flux deviation $ \delta\Phi$. We provide estimates for the set of parameters optimized for
the observation of $4e$-supercurrent.
\end{abstract}

\maketitle

\section{Introduction}

Pairing of Cooper pairs  in frustrated Josephson junction arrays
was theoretically proposed
recently~\cite{Doucot,IoffeFeigelman02,Doucot03} in the search  of
topologically protected nontrivial quantum liquid states. The
simplest system where such a phenomenon could be observed was
proposed by Doucot and Vidal in~\cite{Doucot}. It consists of a
chain of rhombi (each of them being small ring of 4
superconductive islands connected by 4 Josephson junctions) placed
into transverse magnetic field. cf. Fig. ~1. It was shown
in~\cite{Doucot} that in the fully frustrated case (i.e. magnetic
flux through each rhombus $\Phi = \frac12\Phi_0 = \frac{hc}{4e} $)
usual tunnelling of Cooper pairs along the chain is blocked due to
destructive inteference of tunneling going through two paths
within the same rhombus, while  correlated 2-Cooper-pair transport
survives. Evidently, experimental observation of such a phenomenon
(detected as anomalous period $\frac12\Phi_0 $ of the global
supercurrent along the chain) would be very desirable. However,
theoretical results of Ref.~\cite{Doucot} refer to the situation
when Coulomb energy is determined by self-capacitances $C_0$ of
individual superconductive islands, whereas in real submicron
Josephson-junction arrays  capacitances of {\em junctions} $C$
 dominate, cf. e.g.~\cite{capacitances}.
In this paper we reconsider the model of Ref.~\cite{Doucot} for the
experimentally relevant situation $C \gg C_0$. This case is also simpler for theoretical
treatment, since Lagragian of the system becomes a sum of terms, such that each of them belongs
to individual rhombus only. The only source of coupling between different rhombi is  the  periodic
boundary condition along the chain. The method to treat similar
 problem was developed recently by Matveev, Larkin and Glazman~\cite{Larkin} (MLG). They  considered
simple chain of $N$ Josephson junctions in the closed-ring
geometry, and reduced calculation of supercurrent in large-$N$
limit to  the solution of a Schr\"{o}edinger equation for a
particle moving in a periodic potential $\sim \cos{x}$, with
appropriate boundary condition. MLG assumed (we will do the same)
that Josephson energy $E_J$ of junctions is large compared to
their charging energy $E_C=e^2/2C$.
 We will generalize the MLG method in order to use it for
the case of ring of rhombi. It will be shown that in our case
fictitios particle of the MLG theory is still moving in the
$\cos$-like potential, but it acquires now large {\it spin} $S =
\frac12 N$, where $N$ is the number of rhombi in the ring. In the
maximally frustrated case $ |\Phi_r-\Phi_0/2| \equiv \delta\Phi =0
$ the $x$-projection of the spin is an integral of motion, which
should be chosen to minimize the total energy. As a result, $S_x =
\pm \frac12 N$ and the whole problem reduces to the one studied by
MLG up to trivial redefinition of parameters. In this situation
ground-state energy and supercurrent (which is proportional to
derivative of the ground-state energy over total flux $\Phi_c$)
are periodic function of $\Phi_c$ with period $\Phi_0/2$, i.e.
$4e$-transport takes place. Nonzero flux deviation $\delta\Phi$
produces longitudinal field $h_z$ coupled to the $z$-component of
spin of fictitios particle, which acquires now nontrivial
dynamics. We show that in the limit of sufficiently long rhombi
chain the whole problem can be analyzed in terms of semiclassical
dynamics of a particle with a large spin under spin-dependent
potential barrier. In general,  there are two  tunnelling
trajectories, one of them corresponds to usual $2e$ transport,
whereas another to $4e$ transport.  Comparing actions of these
trajectories for different $\delta\Phi$, we find critical flux
deflection $\delta\Phi_c$ as function of the ratio $E_J/E_C \gg
1$.

The rest of the paper is organized as follows: in Sec. II we
define our model and identify its classical states; in Sec. III we
derive effective Hamiltonian which governs quantum phase slip
processes and calculate supercurrent as function of the flux
deflection  $\delta\Phi_c$; in Sec. IV we consider current-bias
chain with $I > I_c$ and calculate voltage $V(I)$ via the rate of
incoherent quantum phase slips. Our conclusions and suggestions
for the experiment are presented in Sec. V.  Finally, in Appendix
A somewhat tedious calculation of current-phase relation is
presented.

\section{The model and its classical states}

\begin{figure}[h] \centering
\includegraphics[width=450pt]{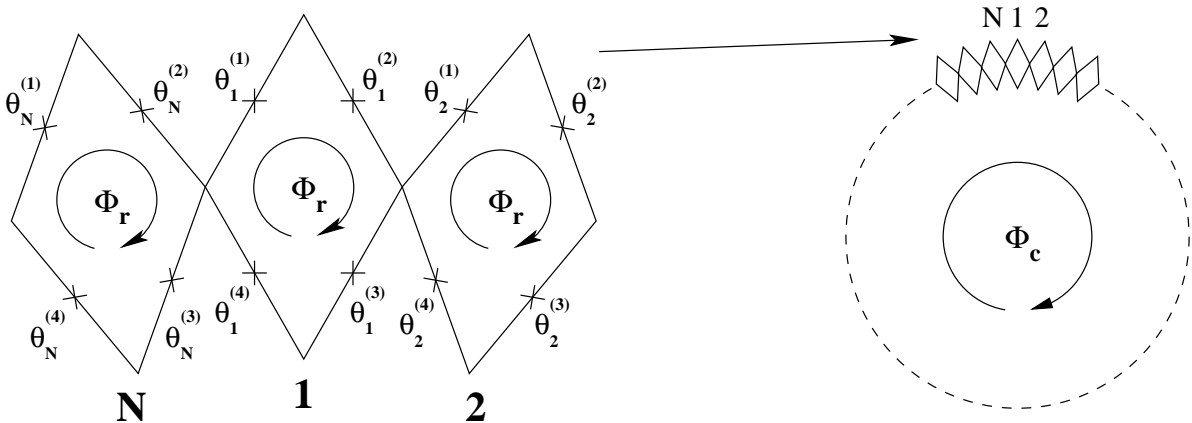}
 \caption{\small The chain of rhombi as a closed ring.}
\label{chain}
\end{figure}

We study a chain of $N$ rhombi shown in Fig.~\ref{chain}.
Each rhombi consists of four superconductive islands
connected by tunnel
junctions with Josephson coupling energy $E_J=\hbar I_c^0/2e$;
charging energy $E_C$ is determined by capacitance  $C$ of junctions,
$E_C = e^2/2C$ (we neglect self-capacitances of islands which are assumed to be much smaller
than $C$).  Below we consider Josephson current along the chain of $N \gg 1$ rhombi
and assume that the chain is  of the ring shape, with total magnetic
flux $\Phi_c$ inside the ring.
 We also denote by $\Phi_r$ the flux per elementary rhombus
plaquette and define phases $\gamma$ and $\varphi$:
\begin{equation}
\gamma=2\pi \frac{\Phi_c}{\Phi_0}\,, \qquad
\varphi=2\pi\frac{\Phi_r}{\Phi_0}\,,
\end{equation}
where $\Phi_0=h/2e$ is the superconducting flux quantum. We study
the situation than $\Phi$ is close to $\Phi_0/2$, i.e.
$\delta=\varphi-\pi\ll 1$.

Assuming that the charge transport through the system at $\delta=0$ is
carried out by charges $4e$,  we expect that in this case
dependence of the current in the chain on the external flux
$\Phi_c$ is periodic with period $\Phi_0/2$. Below we calculate
the $\Phi_0/2$-periodic current at $\delta=0$. We also show that
at small $\delta$ the current through the system has two
components $I_{4e}$ and $I_{2e}$ with periods $\Phi_0/2$ and
$\Phi_0$ respectively. The first component corresponds  to the
current carried by pairs of Cooper pairs and the second one corresponds to
the single Cooper pair transport. At very small $\delta$ the
current $I_{4e}$ dominates over $I_{2e}$. We will refer to this
regime as to $4e$-regime. At large enough $\delta$ the opposite situation
($2e$-regime) is realized. We will find below the
crossover point $\delta_c$ between these two regimes.

The system  is described by the following
imaginary-time action
\begin{equation}
S_E=\int
dt\sum_{n=1}^N\sum_{m=1}^4\left\{\frac{1}{16E_C}\left(\frac{d\theta^{(m)}_n}{dt}
\right)^2-E_J\cos\theta^{(m)}_n\right\}. \label{basic_action}
\end{equation}
Here the variable $\theta^{(m)}_n$ is the phase difference across
the $m$-th junction in the $n$-th rhombus (see Fig. 1).
Taking into account that each rhombus is pierced by flux
$\Phi_r$ and the flux through the whole chain is $\Phi_c$ we derive
the following additional conditions
\begin{equation}
\sum_{m=1}^4\theta^{(m)}_n=\varphi\,, \qquad n=1,2,..,N\,,
\label{varphi}
\end{equation}
\begin{equation}
\sum_{n=1}^N\left(-\theta^{(3)}_n -
\theta^{(4)}_n\right)=\gamma\,. \label{gamma}
\end{equation}

In this paper  we consider the case of strong coupling between
grains $E_J\gg E_C$. This enables us to use semi-classical
approximation for calculating the energy spectrum of the system.
At $E_C=0$ the phases $\theta^{(m)}_n$ become classical variables
and the energy states of the chain can be found by minimizing the
sum of Josephson energy terms in action~(\ref{basic_action}). Let
us introduce variables $\theta_n=-\theta^{(3)}_n-\theta^{(4)}_n$,
where $\theta_n$ is the phase difference along the diagonal
 of the $n$-th rhombus.  It is convenient to make minimization in two steps.
First of all the Josephson energy of a single rhombus under the
fixed flux through the rhombus and under the fixed phase
difference $\theta_n$ is minimized. For the Josephson energy of
the chain we then get for $\delta \ll 1$:
\begin{equation}
E=-2\sqrt{2}E_J\sum_{n=1}^N\left(1+\frac{1}{4}\delta\sigma_n^z\right)
\cos\left(\frac{\theta_n}{2}-\beta_n\right)\,,
\label{E_under_fixed_theta_n}
\end{equation}
\begin{equation}
\sin\beta_n=\pm\frac{\sigma^z_n}{\sqrt{2}}
\left(1-\frac{\delta}{4}\sigma_n^z\right)\,,\qquad
\cos\beta_n=\pm\frac{1}{\sqrt{2}}\left(1+\frac{\delta}{4}\sigma_n^z\right)\,.
\label{beta_n}
\end{equation}
Plus and minus signs in (\ref{beta_n}) correspond to positive (resp. negative)
values of  $\cos\frac{\theta}{2}$. Here
we have introduced an important notation $\sigma^z_n={\rm sign}\,
\sin\theta_n$. It can be easily shown that at $\delta=0$ each
individual rhombus has two classical ground states with equal
energies. This states differ only in the sign of the
superconducting current circulating around the plaquette which
corresponds to the binary variable $\sigma_n^z$.

Now we have to minimize the energy (\ref{E_under_fixed_theta_n}) with respect
to  phases $\theta_n^{(m)}$  subject to the constrains (\ref{gamma}).
Assuming $\delta$ to be small and $N$ to be large we get
\begin{equation} E_{m, \sigma}\approx
\frac{E_J\sqrt2}{4N}(\widetilde{\gamma}-\pi N/2-\pi S^z- 2\pi m)^2
-\sqrt{2}\delta S^z E_J+{\rm Const}.
\label{E_under_flux_dif_from_pi}
\end{equation}
Here $m$ is an arbitrary integer (which has the same meaning
as in the MLG paper~\cite{Larkin}) and
\begin{equation}
s_n^z=\frac{1}{2}\sigma_n^z\,, \qquad
S^z=\frac{1}{2}\sum_{n=1}^N\sigma_n^z\,, \qquad
\widetilde{\gamma}=\gamma+\frac{N\varphi}{2}=\gamma+\frac{\pi
N}{2}+\frac{N\delta}{2}\,.
\end{equation}
In the above equation $s_n^z$  can be considered as $z$-projection
of the "spin" $\frac12$ which describes  binary degeneracy of
states of the $n$-th rhombi. Then $S_z$ corresponds to the
$z$-projection of the total large spin $\bf S$ describing the
whole  rhombi chain. For clarity everywhere in this paper we will
refer to the case of even number of rhombi. Then total spin $S$
and eigenvalues of its projection $S^z$ are integer. Classical
states of the chain are characterized by individual spin
projections $\sigma_n^z$ for each rhombus, and by  collective
integer-valued variable $m$. We will denote these states by
$\left|m,\{\sigma_n^z\}\right>$ or $|m,\sigma>$. Physically,
classical state of the chain is characterized by the global
current $I$  along the chain, and by the signs of local currents
flowing in each of $N$  rhombi.  Nonzero charging energy $E_C$
provides quantum phase slips  in each of $4N$  Josephson junction;
these  processes mix different classical states leading to
formation of the fully quantum ground state. Below we derive
effective
 Hamiltonian acting on the space of classical states, and find ground-state energy
 $E_0(\gamma)$ and corresponding supercurrent.

\section{Quantum fluctuations of rhombi and supercurrent}
\begin{figure}
\includegraphics[width=300pt]{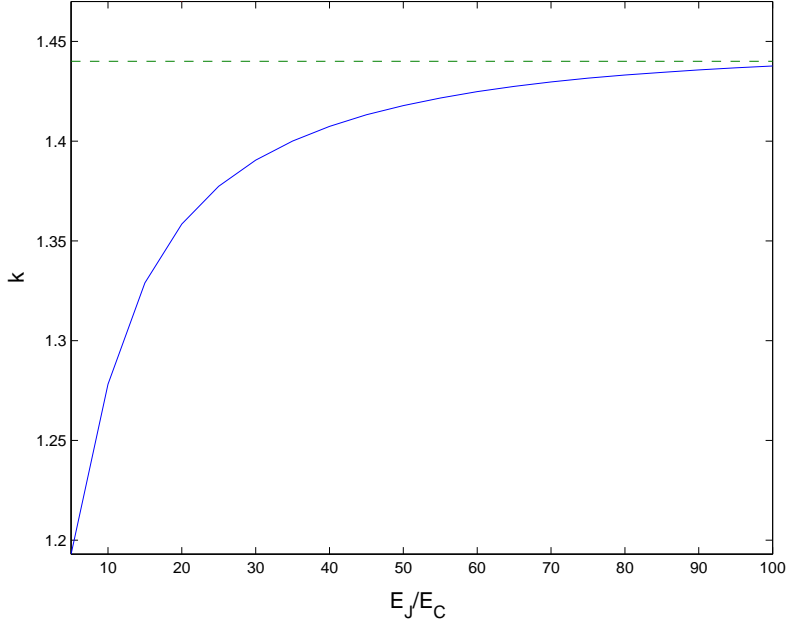}
\caption{\small The numerical factor $k$ as a function of $E_j/E_c$.}
\label{k_fig}
\end{figure}

We  turn to analysis of  quantum fluctuations
of $\theta_n^{(m)}$ at finite $E_C$. The most important type of
these fluctuations involve an instanton (quantum phase slip, QPS), i.e.
a trajectory that begin at one of the
minima~(\ref{E_under_flux_dif_from_pi}) of the potential energy in
action~(\ref{basic_action}) at $t=-\infty$ and ends at another
minima at $t=+\infty$.  There are two kinds of trajectories: the first
starts at $\left|m, \{\sigma_n^z\}\right>$ and ends at
$\left|m,\{\sigma_n^z+2\delta_{nk}\}\right>$ for arbitrary $1\leq
k \leq N$, $\sigma^z_k=-1$, whereas the second start at
$\left|m, \{\sigma_n^z\}\right>$ and end at
$\left|m+1,\{\sigma_n^z-2\delta_{nk}\}\right>$ for arbitrary
$1\leq k \leq N$, $\sigma^z_k=1$. Any trajectory of the first kind
corresponds to QPS in $\theta_k^{(1)}$ or $\theta_k^{(2)}$, whereas
trajectory of the second kind corresponds to QPS
$\theta_k^{(3)}$ or $\theta_k^{(4)}$. Note that at $\delta=0$ and
$\gamma=\pi/2$ all these trajectories starting at $\left|m,
\{\sigma_n^z\}\right>$ with $2m+S^z=0$ connect the minima with
equal energies. Thus they are  important for restoring
symmetry of the system which is classically broken. Let us denote as
$\upsilon$ the  amplitude of a QPS in one contact.
At large $N \gg 1$ this amplitude does not differ from the "spin flip"
 amplitude for  a single rhombus at $\Phi_r \approx \Phi_0/2$.
 In this approximation we can use result from Ref.~\cite{IoffeFeigelman02}:
\begin{equation}
\upsilon\approx k
\left(E_J^3E_C\right)^{1/4}\exp\left(-1.61\sqrt{\frac{E_J}{E_C}}\right).
\label{upsilon_gen}
\end{equation}
where $k$ is a numerical factor of the order of unity. Comparison
with direct numerical diagonalization~\cite{Ioffe_sp} of of
low-lying spectrum of a single frustrated  rhombi shows that
coefficient $k$ grows from approximately 1.3 to 1.44 as the ratio
$E_J/E_C$ varies from $10$ to infinity, cf. Fig.~\ref{k_fig}. The
instantons account for the possibility of the system tunneling
between different minima~(\ref{E_under_flux_dif_from_pi}) of the
potential energy in action~(\ref{basic_action}). The effect of the
instantons on the ground state energy can be represented  by
 a tight-binding Hamiltonian defined as
\begin{multline}
\widehat{H}\left|m,\{ \sigma_n^z\}\right>=
E_{m,\sigma}\left|m,\{\sigma_n^z\}\right> +\\ 2\upsilon
\sum_{k=1}^N \left|m,\{\sigma_1^z\,,\cdots\sigma_{k-1}^z\,,
-\sigma_k^z\,, \cdots\sigma_{k+1}^z\,,\sigma_N^z\}\right> +
\\2\upsilon \sum_{k=1\,,\; \sigma_k^z=1}^N
\left|m+1,\{\sigma_1^z\,,\cdots\sigma_{k-1}^z\,, -\sigma_k^z\,,
\cdots\sigma_{k+1}^z\,,\sigma_N^z\}\right> +\\
2 \upsilon \sum_{k=1\,,\; \sigma_k^z=-1}^N
\left|m-1,\{\sigma_1^z\,,\cdots\sigma_{k-1}^z\,, -\sigma_k^z\,,
\cdots\sigma_{k+1}^z\,,\sigma_N^z\}\right> \,.
\label{H_acting_over_m_s}
\end{multline}

To find the ground state energy $E(\gamma)$ it is convenient to
make Fourier transformation over variable $m$ according to
\begin{equation}
\left|x,\sigma\right>
=\sum_m\exp\left\{2i\left(2m-\frac{\widetilde{\gamma}}{\pi}+S^z+\frac{N}{2}\right)x\right\}
\left|m,\sigma\right>, \quad S^z=\sum_{n=1}^N s_n^z. \label{Fur'e}
\end {equation}
Note that not all vectors of our new basis~(\ref{Fur'e}) are independent. It is
easy to see from~(\ref{Fur'e}) that
$\left|x+\pi/2, \sigma\right>=e^{-i\widetilde{\gamma}+i\pi S^z+i\pi N/2}
\left|x,\sigma\right>$.
Considering any system's state of the form
$\left|\psi\right>=\sum_{x,\sigma}\psi(x,\sigma)\left|x,\sigma\right>$
we should impose on the wavefunction $\psi(x, \sigma)$ a twisted boundary condition
\begin{equation}
e^{i\pi\widehat{S}^z+i\pi N/2}\psi\left(x+\pi/2,\sigma\right)
=e^{i\widetilde{\gamma}}\psi(x,\sigma)\,. \label{boundary}
\end{equation}
Here we have introduced operator $\widehat{S}^z$
acting on the spin variables of the system according to standart rules.

 The resulting Schr\"{o}dinger equation acquires the form
\begin{equation}
\frac{\partial^2\psi}{\partial x^2} +(b- 2 w \cos 2x\cdot
\widehat{S}^x+2h\widehat{S}^z) \psi=0\,,
\label{main_res_under_flux_dif_from_pi}
\end{equation}
where
\begin{equation}
b=\frac{16NE}{\sqrt2 E_J \pi^2}\,, \qquad
w=\frac{64N\upsilon}{\sqrt2 E_J\pi^2}\,,\qquad
h=\frac{8N\delta}{\pi^2}\,. \label{bwh}
\end{equation}

Note that  symmetry group of the Hamiltonian corresponding
to the equation~(\ref{main_res_under_flux_dif_from_pi}) includes
 transformations $U_n=e^{i\pi n \left(\widehat{S}^z+N/2\right)}
\widehat{T}_{\pi n/2}$. Here $\widehat{T}_a$ is operator of the translation
over distance $a$ along the $x$-axis.  Equation~(\ref{boundary}) shows that the
parameter $\widetilde{\gamma}$ specifies different irreducible
representations of the  symmetry group.

The
equations~(\ref{main_res_under_flux_dif_from_pi},\ref{boundary})
allow comprehensive analytical investigation for the case when the
flux per single rhombus  equals $\Phi_0/2$.
For such a system $h=0$ and the Hamiltonian commutes with
$S^x$. However, variables $x$ and $S$ are still coupled due to boundary condition
~(\ref{boundary}). Therefore we look for wavefunction in the form
\begin{equation}
\psi(x,\sigma)=e^{i\widetilde{\gamma}
}\left|S^x\right>\phi(x)+e^{i\pi
\widehat{S}^z+i\pi N/2}\left|S^x\right>\phi(x+\pi/2)
\label{wave_func}
\end{equation}

Then for $\phi(x)$ we get standard Mathieu equation:
\begin{equation}
\frac{\partial^2\phi}{\partial x^2} +(b- 2 q\cos 2x) \phi=0\,,
\label{main_res_for_delta_eq_zero}
\end{equation}
where $q=wS^x$. Boundary condition~(\ref{boundary}) now reads  as follows:
\begin{equation}
\phi(x+\pi)=e^{2i\widetilde{\gamma}}\phi(x)\,.
\label{x_dependent_boundary}
\end{equation}
The  ground state of the system defined by Eqs. (\ref{main_res_for_delta_eq_zero},
\ref{x_dependent_boundary}) corresponds to  maximal absolute value of $S_x$,
equal to $N/2$.
In other words, in the ground state all rhombi in
the chain are either in symmetric superpositions or in antisymmetric superpositions,
of their double-degenerate classical states. Thus there are two degenerate
eigenstates
\begin{equation}
|0^+_{\widetilde{\gamma}}\rangle=\phi_{\widetilde{\gamma}}(x)|S^x=N/2\rangle
\qquad {\rm and }\qquad
|0^-_{\widetilde{\gamma}}\rangle=\phi_{\widetilde{\gamma}}(x+\pi/2)|S^x=-N/2\rangle =
\widehat{U}_1 |0^+_{\widetilde{\gamma}}\rangle\,,
\end{equation}
with the same lowest energy $E_0$. This degeneracy is a direct
consequence of the fact that for $h=0$ (fully frustrated chain)
the Hamiltonian has two symmetry operators $S^x$ and $U_1$ which
do not commute with each other (thus double-degeneracy refers to
all states, not only to the ground-state). Eigenstates
$|0^\pm_{\widetilde{\gamma}}\rangle$ constitute a basis where
$\widehat{S}^x$ operator is diagonal. Coming back to the original
problem defined by Eqs.(\ref{boundary},
\ref{main_res_under_flux_dif_from_pi} ) we note, that in
accordance with~(\ref{wave_func}) the correct (unique) eigenstate
obeying Eq.(\ref{boundary}) can be constructed as specific linear
combination of  $|0^+_{\widetilde{\gamma}}\rangle$ and
$|0^-_{\widetilde{\gamma}}\rangle$:
\begin{equation}
|G_{\widetilde{\gamma}}\rangle =
\frac{e^{i\widetilde{\gamma}}|0^+_{\widetilde{\gamma}}\rangle+
|0^-_{\widetilde{\gamma}}\rangle}{\sqrt{2}}
\label{G}
\end{equation}
which diagonalize operator $U_1$. The eigenstate
$|G_{\widetilde{\gamma}}\rangle$ is similar to the eigenstates $|G\rangle$ of
Ref.~\cite{IoffeFeigelman02}, cf. Eq.(5) of that paper.

 It follows from the
boundary condition~(\ref{x_dependent_boundary}) that
shift of the phase $\widetilde{\gamma}$ by $\pi$ does not change the boundary
problem  defined by Eqs. (\ref{main_res_for_delta_eq_zero},
\ref{x_dependent_boundary}).
Thus  the ground-state energy of the system and the
supercurrent through the circuit are periodic functions of the flux
$\Phi_c$  with  period $\Phi_0/2$.

At $Nw\sim N^2\upsilon/E_J\ll 1$  fluctuations are weak, the amplitude of
potential energy in~(\ref{main_res_for_delta_eq_zero}) is small and its
effect is most significant when $4\Phi_c/\Phi_0$ is integer and
the energy levels $E_{m,\sigma}$ are degenerate. In this regime
the usual approximation for semiclassical weak link is valid, and for the persistent
current through the circuit we obtain
\begin{equation}
I(\widetilde{\gamma})=\frac{2e}{\hbar}\frac{dE}{d\gamma}= {\rm
sign}\,\widetilde{\gamma}\frac{\sqrt{2}I_c^0\pi}{4N}
\left(1-\frac{2|\widetilde{\gamma}|}{\pi}\right)
\left(\frac{1}{\sqrt{\left(1-2|\widetilde{\gamma}|/\pi
\right)^2+(q/2)^2}}-1\right)\,. \label{weak_fl}
\end{equation}
Phase-dependent current $I(\widetilde{\gamma})$ is described by
equation~(\ref{weak_fl}) for $-\pi/2<\widetilde{\gamma}<\pi/2$ and is a
periodic function of $\widetilde{\gamma}$ with period $\pi$. So in the regime
of weak fluctuations the dependence $I(\widetilde{\gamma})$ demonstrates
sawtooth behavior slightly rounded due to fluctuations.

The opposite limit $Nw\gg1$ corresponds to the regime of strong
fluctuations. In this case the dependence of the eigenvalue $b$ on
the phase $\widetilde{\gamma}$ is exponentially weak~\cite{Abramowitz}:
\begin{equation}
b=-2q+16\sqrt{\frac{2}{\pi}}q^{3/4}e^{-4\sqrt{q}}(1-\cos2\widetilde{\gamma})
\label{bqphi}
\end{equation}
and for the persistent current in the ground state we find
\begin{equation}
I(\widetilde{\gamma}) =32\cdot2^{3/8}I_c^0\left(\upsilon/E_J\right)^{3/4}\sqrt{N}
\exp\left\{{-\frac{16\cdot2^{1/4}}{\pi}N\sqrt{\frac{\upsilon}{E_J}}}\right\}
\sin2\widetilde{\gamma}\,. \label{I(0.5)}
\end{equation}
Equation (\ref{I(0.5)}), together with Eq.(\ref{upsilon_gen}),
presents one of our main results: it gives the
amplitude of $4e$ - periodic Josepshson current in the regime
of maximal frustration.

Let us now turn to the investigation of the general situation
described by equations~(\ref{main_res_under_flux_dif_from_pi})
and~(\ref{boundary}). As was mentioned above if the flux per
elementary plaquette differs slightly  from half supercondicting
flux quantum the persistent current through the chain has two
components $I_{4e}$ and $I_{2e}$. In the regime of strong fluctuations
both these currents are exponentially small. The main exponential
factors in the expressions for them can be found on the basis of
equation~(\ref{main_res_under_flux_dif_from_pi}) using the
semi-classical approximation.

Note that equation~(\ref{main_res_under_flux_dif_from_pi})
corresponds to a particle of mass $1$ with spin $S$ moving in
one-dimensional potential
\begin{equation}
U(x,\vec{S})=w \cos 2x\cdot S^x-hS^z\,,
\label{basic_potential_for_spin_system}
\end{equation}
the particle energy being $E^0=b/2$. So denoting by $\theta$ and
$\phi$ the angels determining the spin direction, we can write the
imaginary time tunneling amplitude in the form of path
integral~\cite{Klauder}
\begin{multline}
\left<\theta_2\,,\phi_2\,,x_2\right|e^{-T\widehat{H}}\left|\theta_1\,,\phi_2\,,x_1\right>=\\
\int_{\theta_1,\,\phi_1,\,x_1}^{\theta_2,\,\phi_2,\,x_2}
\mathcal{D}\Omega\mathcal{D}x
\exp\left\{-\int_{-T/2}^{T/2}d\tau\left(i
S(1-\cos\theta)\dot{\phi}
+\frac{\dot{x}^2}{2}+U(x\,,\vec{S})\right)\right\}
\label{basic_path_integral_for_spin_system}
\end{multline}
For our future purposes it is more convenient to use the above
path integral in another form also derived in~\cite{Klauder}:
\begin{multline}
\left<\vec{S}_2\,,x_2\right|e^{-T\widehat{H}}\left|\vec{S}_1\,,x_1\right>=\\
\int_{\vec{S}_1,\,x_1}^{\vec{S}_2,\,x_2}
\mathcal{D}\vec{S}\mathcal{D}x\,\delta\left(\vec{S}^2-S^2\right)
\exp\left\{-\int_{-T/2}^{T/2}d\tau\left(i\frac{S^x\dot{S}^y-\dot{S}^xS^y}{S+S^z}
+\frac{\dot{x}^2}{2}+U(x\,,\vec{S})\right)\right\}
\label{basic_path_integral_for_spin_system_for_lin}
\end{multline}

We will analyse the
expression~(\ref{basic_path_integral_for_spin_system_for_lin}) for
the limit of relatively large $\delta$ when $h\gg w$. In this regime
the field $h$ in~(\ref{basic_potential_for_spin_system}) fixes the
direction of $\vec{S}$ so that $S^x$ and $S^y$ are always small.
Therefore we can linearize the action
in~(\ref{basic_path_integral_for_spin_system}) with respect to
$S^x$ and $S^y$. After linearization we can easily exclude the
variable $S^y$ using the equations of motion.
The substitution of variables
 $S^x\longrightarrow\sqrt{Sh}y$, $\tau\longrightarrow
\tau/h$ leads to the path integral
\begin{equation}
\left<y_2\,,x_2\right|e^{-T\widehat{H}}\left|y_1\,,x_1\right>=
\int_{y_1,\,x_1}^{y_2,\,x_2}
\mathcal{D}x\mathcal{D}y\, exp\left(-S_E\right)
\label{2D_path_integral}
\end{equation}
where the action
\begin{equation}
S_E=\frac{h}{2}\int_{-\infty}^{+\infty}d\tau\left\{\dot{x}^2+\dot{y}^2+
U_{eff}(x,y)\right\}\,, \label{linearized_action}
\end{equation}
\begin{equation}
U_{eff}(x,y)=(y+d\cos{2x})^2+d^2\sin^2{2x}\,,\qquad
d=\sqrt{\frac{w^2S}{h^3}}=\frac{\sqrt{2}\pi}{\delta^{3/2}}\frac{\upsilon}{E_J}\,.
\label{effektive_potential}
\end{equation}
The appropriate equations of motion are
\begin{equation}
\ddot{x} +2dy\sin{2x}=0\,, \label{equation_for_x_aft_subst}
\end{equation}
\begin{equation}
\ddot{y}-y-d\cos{2x}=0\,. \label{equation_for_y_aft_subst}
\end{equation}

Using semi-classical approximation we should first determine the
classical minima of the potential~(\ref{effektive_potential}).
Within the same limit $h\gg w$ we find that $U_{eff}$ has two
groups of minima (we call them "even" and "odd" minima)
\begin{equation} x=\pi n\,,\qquad
y=-d\,,\label{even}
\end{equation}
\begin{equation} x=\frac{\pi}{2}+\pi n\,,\qquad
y=d\,,\label{odd}
\end{equation}
where $n$ is an arbitrary integer. All these minima correspond to
the same value of $U_{eff}=0$. So we have to consider two types of
tunnelling trajectories.
Trajectories of the first type connect minima of the same group,
i.e. "even-even" and "odd-odd", and corresponding variation of the variable
$x$ between minima is $\pm \pi$, whereas $y$ returns to its original value.
Trajectories of the second type connect minima of opposite parity (i.e. opposite
signs of $y$), and change $x$ variable by $\pm \frac{\pi}2 $. It is not difficult to
see from Eqs.(\ref{Fur'e},\ref{main_res_under_flux_dif_from_pi},\ref{boundary} ),
 that increment $\Delta x$ of the variable $x$ along tunnelling trajectory is in one-to-one
correspondence to the elementary charge transported along the rhombi chain:
$q_0 = \frac{4e}{\pi}\Delta x$. Therefore trajectories of the first type lead to
$4e$ - supercurrent,
 whereas trajectories of the second type produces usual $2e$-quantized
supercurrent. The amplitudes of the supercurrent components are determined
(cf. Appendix for the direct derivation)
primarily by the classical actions on corresponding trajectories:
\begin{equation}
I(\gamma) = I_{2e}\sin\widetilde{\gamma} + I_{4e}\sin(2\widetilde{\gamma}),
\label{gen_cur}
\end{equation} where
\begin{equation}
I_{4e} = A_{4e}  \exp(-S_E^{4e}),  \quad  I_{2e} = A_{2e} \exp(-S_E^{2e}) \, ,
\label{components}
\end{equation}
and  $S_E^{4e}$ and $S_E^{2e}$ are the values of tunnelling actions on
trajectories of the first and second type
respectively. Both $S_E^{4e}$ and $S_E^{2e}$  are large in the
region of strong fluctuations $Nw \gg 1$, thus the total
supercurrent will  in general be dominated by least-action processes.

To compare actions $S_E^{4e}$ and $S_E^{2e}$ we note that the
dynamical  system~(\ref{equation_for_x_aft_subst})
and~(\ref{equation_for_y_aft_subst}) has two characteristic
frequencies. The first one characterises   "spin" \, subsystem
with $\omega_s=1$, whereas the second one is the frequency of the
"x" \, subsystem, $\omega_x\sim d$, since
 typical value of $y$ in~(\ref{equation_for_x_aft_subst})
is~$d$. Therefore at  $d \ll 1$ - i.e. at sufficiently large flux deflections
$\delta$, the "spin variable" \, $y$ is fast and can be integrated out
in adiabatic approximation, which leads to
\begin{equation}
S_E=h\int d\tau\left\{\frac{\dot{x}^2}{2}-\frac{d^2}{4}\cos 4x\right\}\,,
\label{small_d_action_with_x}
\end{equation}
\begin{equation}
S_E^{4e}\approx 2hd, \quad  {\rm and} \quad S_E^{2e}\approx hd, \,\,\, {\rm at}
\quad d\ll 1
\label{smalld}
\end{equation}
The dominant process is thus usual $2e$ transfer. Comparing the
action~(\ref{small_d_action_with_x}) with the action corresponding to the
Shr\"{o}dinger equation~(\ref{main_res_for_delta_eq_zero})
\begin{equation}
S_E^0=\int d\tau \left\{\frac{\dot{x}^2}{2}+q\cos 2x\right\}
\label{S_E^0}
\end{equation}
and using~(\ref{bqphi}) we obtain supercurrent amplitude
\begin{equation}
I_{2e} \approx 32\cdot2^{1/4}I_c^0\sqrt{N}
\left(\frac{\upsilon}{E_J\sqrt{\delta}}\right)^{3/2}
\exp\left\{-
\frac{8\sqrt{2}}{\pi}\frac{N\upsilon}{E_J\sqrt{\delta}}\right\}
\quad {\rm at} \quad \frac{1}{\delta^{3/2}}\frac{\upsilon}{E_J}
\ll 1 \label{I2e}
\end{equation}
At small flux deflection $\delta$  the parameter $d \gg 1$ and the spin variable
$y$ is relatively slow and almost does not change on the type-1 trajectory. The dominant
trajectory is then $4e$-one. Assuming $y$ to be constant,  we get
\begin{equation}
S_E^{4e}=h\int d\tau\left\{\frac{\dot{x}^2}{2}-d^2\cos2x\right\}=4hd
\label{large_d_action_with_x}
\end{equation}
Taking into account also the first-order term of
perturbation theory over $1/d \ll 1$, we find $S_E^{4e} = h(4d-1)$.
 Comparison of the action~(\ref{large_d_action_with_x})
with~(\ref{S_E^0}) and~(\ref{bqphi}) allows us to
determine the pre-exponential factor in the expression for the current
\begin{equation}
I_{4e} \approx 128\cdot2^{1/4}I_c^0\sqrt{N}
\left(\frac{\upsilon}{E_J\sqrt{\delta}}\right)^{3/2} \exp\left\{
-\frac{32\sqrt{2}}{\pi}\frac{N\upsilon}{E_J\sqrt{\delta}}
+\frac{8N\delta}{\pi^2}\right\}
 \quad {\rm at} \quad  \frac{1}{\delta^{3/2}}\frac{\upsilon}{E_J}
\gg 1 \label{I4e}
\end{equation}
Note that at $\delta$ determined from the equation $h=w$
(where the linear approximation used to describe the spin
degree of freedom fails), the $4e$-current
from~(\ref{I4e}) matches the exact result for $\delta=0$ presented
in~(\ref{I(0.5)}).

\begin{figure}
\includegraphics[width=300pt]{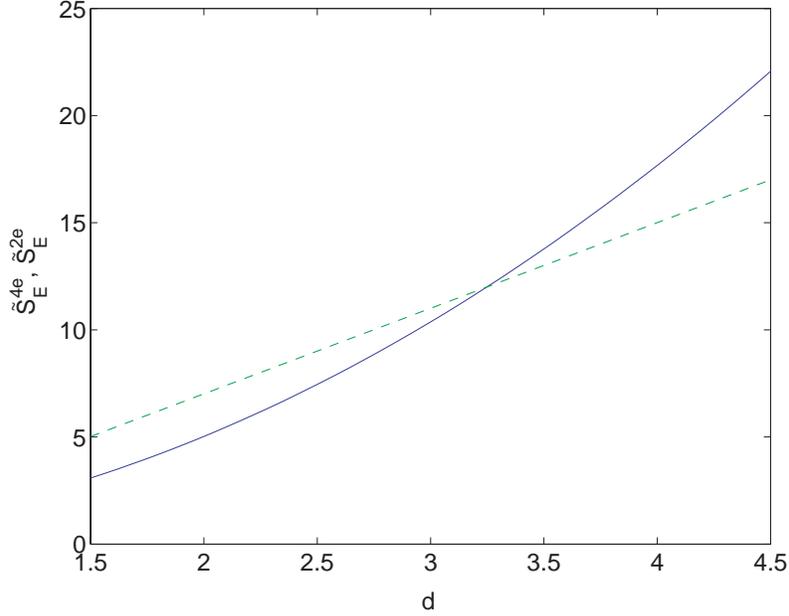}
\caption{\small The results of numerical evaluation of
$\widetilde{S}_E^{2e}(d)$ (solid line) and
$\widetilde{S}_E^{4e}(d)$ (dashed line).} \label{both_actions}
\end{figure}

In the intermediate region of $d \sim 1 $  we analyse equations
(\ref{linearized_action}),~(\ref{equation_for_x_aft_subst})
and~(\ref{equation_for_y_aft_subst})  numerically.
First we write $S_E^{4e}$ and $S_E^{2e}$ in the form
$S_E^{4e}=h\widetilde{S}_E^{4e}(d)$,
$S_E^{2e}=h\widetilde{S}_E^{2e}(d)$. Then  the functions
$\widetilde{S}_E^{4e}(d)$ and $\widetilde{S}_E^{2e}(d)$ depending
on a single parameter $d$ have been evaluated numerically. The
result is presented on Fig.~\ref{both_actions}.
The actions for both types of trajectories are equal
at $d=d_0\approx3.2$, where we have $\widetilde{S}_E^{4e}(d_0) =
\widetilde{S}_E^{2e}(d_0)\approx 11.9$.
Thus the crossover between $4e$-regime and $2e$-regime takes
place at
\begin{equation}
\delta\Phi=\delta\Phi^c=\left(\frac{\upsilon^2}{4\pi
d_0^2E_J^2}\right)^{1/3}\Phi_0\approx
0.2\left(\frac{\upsilon}{E_J}\right)^{2/3}\Phi_0\label{crossover}
\end{equation}
Varying flux $\Phi_r$  in some vicinity of crossover point (\ref{crossover})
one can find both $2e$ and $4e$ components of supercurrent, but their relative
weight is expected to vary strongly with $\Phi_r-\Phi_r^c$,  in some analogy
with phase coexistence near first-order phase transition.

\section{Low-voltage states}

In previous Section we obtained estimates (\ref{weak_fl},\ref{I(0.5)},\ref{I2e},\ref{I4e}) for
 the equilibrium
supercurrent $I(\gamma)$ around the {\it flux-biased} rhombi chain with $N \gg 1$.
Note that the maximum value of this supercurrent is small, compared to
individual critical current of a single junction $I_c^0$, even in the case
of weak quantum fluctuations, cf. Eq.(\ref{weak_fl}). This is due to the
fact that in our analysis we have considered perfectly equilibrium
Josepshson current, which must be $2\pi$-periodic as function of the total
phase bias $\gamma$. Therefore in the long chain phase differences across each
rhombi
scales as $1/N$, leading to $I_c \sim I_c^0/N$ in weak-fluctuation limit $N w \ll 1$
(in the opposite limit $I_c$ is small exponentially in $N$).
It is clear, however, that under the condition of some {\it current bias},
with a fixed $ I \ll I_c^0$,  the chain will be in some
"nearly superconducting" state with a very low voltage,  due to rare
phase slip processes. Below we consider  regime of relatively large
currents (the condition to be specified below), when the processes of
tunnelling in different rhombi are incoherent. In this case mean voltage $V$
along the  whole chain can be estimated just as $N$ times the voltage along
a single rhombus. Below we estimate probability per unit time of an individual
QPS in a single rhombus at the fixed transport current $I \ll I_c^0$, and find
the $V(I)$ dependence.

Introducing variables $\theta=\theta^{(1)}+\theta^{(2)}$,
$\chi_1=\theta^{(1)}-\theta^{(2)}$,
$\chi_2=\theta^{(3)}-\theta^{(4)}$  we can rewrite the
imaginary-time action for a single rhombus carrying external
current $I$ in the form
\begin{equation}
S_E=\int
d\tau\left\{\frac{1}{32E_C}\left(2\dot{\theta}^2+\dot{\chi_1}^2+\dot{\chi_2}^2
\right)+V(\theta,\chi_1,\chi_2)\right\}\,,
\label{fission_action}
\end{equation}
\begin{equation}
V(\theta,\chi_1,\chi_2)=-E_J\left(
2\cos\frac{\theta}{2}\cos\frac{\chi_1}{2}+
2\sin\frac{\theta}{2}\cos\frac{\chi_2}{2}
+\frac{I}{I_c^0}\theta\right)\,.
\label{fission_potential}
\end{equation}
We have assumed here that the flux inside the rhombus equals
half the supercoducting flux quantum.
In order to find the classical states of the rhombus we eliminate $\chi_1$ and $\chi_2$
from~(\ref{fission_potential}) and get
\begin{equation}
V(\theta)=-E_J\left(2\left|\cos\frac{\theta}{2}\right|+
2\left|\sin\frac{\theta}{2}\right|+\frac{I}{I_c^0}\theta\right)\,.
\label{V_of_theta}
\end{equation}
The potential~(\ref{V_of_theta}) has a number of local minima
$\theta_{min}=\theta_0+\pi m$
where $\theta_0$ is determined by the equation
\begin{equation}
\sin\frac{\theta_0}{2}-\cos\frac{\theta_0}{2}=\frac{I}{I_c^0}\,.
\end{equation}
With an appropriate choice of phases $\chi_1$ and $\chi_2$ every
$\theta_{min}$ corresponds to a classical state of the rhombus
localized near this minimum. Due to quantum tunneling  all these
states are metastable  and have finite decay time~$\tau$.

Within the semi-classical approximation (valid for $E_J\gg E_C$)
the decay time $\tau$ is determined by vicinity of a bounce i.e. a
classical trajectory starting at a minimum of the potential
energy~(\ref{fission_potential}) coming close to another one and
then going back to the first minimum~\cite{Coleman,
Larkin_Ovchinnikov}. To be specific we will refer here to the
decay rate of a state corresponding to $\chi_1=0$, $\chi_2=0$ and
$\theta=\theta_0$. Decay of this state goes via one of two
possible bounce trajectories (for $I>0$). One of them passes near
$\theta=\theta_0$, $\chi_1=2\pi$ and $\chi_2=0$ while the other
passes near $\theta=\theta_0$, $\chi_1=-2\pi$ and $\chi_2=0$. Both
these bounces give equal contribution to the  decay rate.

Let us denote  by $q=(\theta, \chi_1, \chi_2)^T$ --- the
three-dimensional column-vector in the coordinate space of the
rhombus. We also introduce $q_0(\tau)=(\theta_0, 0, 0)^T$  as the
trajectory corresponding to the system being at the minimum of the
potential~(\ref{fission_potential}) and $q_b(\tau)$ as the bounce
trajectory which can be determined by solving the classical
equations of motion. The decay probability per unit time of the
unstable state is given by~\cite{Coleman,Larkin_Ovchinnikov}
\begin{equation}
1/\tau=2\left(\frac{S_E[q_b]}{2\pi}\right)^{1/2}e^{-S_E[q_b]}
\left|Det'\left(\frac{\delta^2S_E}{\delta q^2}\right)_{q=q_b}\right|^{-1/2}
\left|Det\left(\frac{\delta^2S_E}{\delta q^2}\right)_{q=q_0}\right|^{1/2}
\end{equation}
where $Det'$ indicates that the zero eigenvalue is to be omitted when computing the
determinant.

\begin{figure}[t]
\includegraphics[width=300pt]{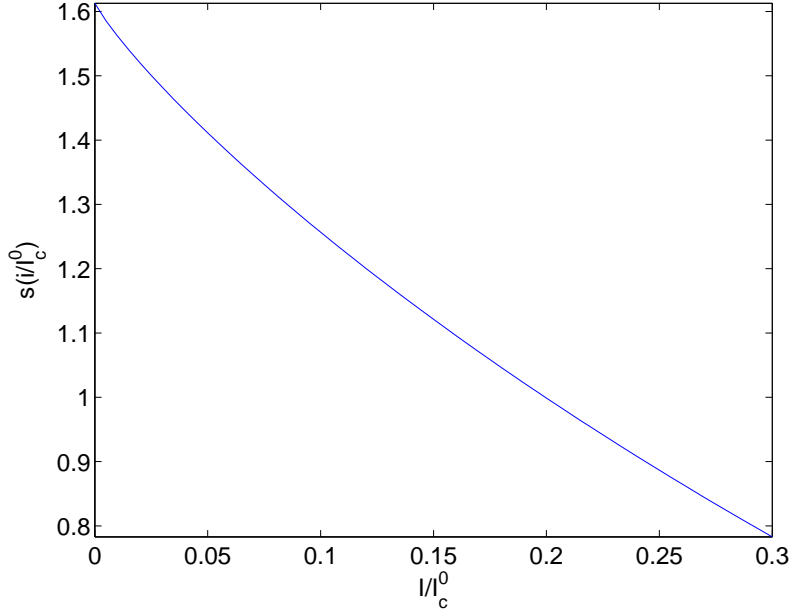}
\caption{\small Results of numerical evaluation of $s(I/I_c^0)$.}
\label{action_fig}
\end{figure}

After changing the time scale according to  $\tau\longrightarrow
\tau/\sqrt{E_JE_C}$ the bounce action can be rewritten as
$S_E[q_b]=2\sqrt{E_J/E_C}\,s(I/I_c^0)$ and for the inverse decay
time we obtain
\begin{equation}
\frac{1}{\tau}\approx2\frac{(E_J^3E_C)^{1/4}}{\hbar}
K(I/I_c^0)\exp\left(-2\sqrt{\frac{E_J}{E_C}}\,s(I/I_c^0)\right)\,.
\end{equation}
Here $K(I/I_c^0)$ is a numerical factor of order one. The function $s(I/I_c^0)$
depending on the only parameter $I/I_c^0$ can be evaluated numerically
by solving the Lagrangian equations
for the action~(\ref{fission_action}) with the appropriate
boundary conditions. The result is presented on Fig.~\ref{action_fig}.
Let us assume that the current $I$ is not very small so that the
energy difference $\delta V= \pi E_JI/I_c^0$ between two nearest
minima of the potential~(\ref{fission_potential}) is much larger
than the  quantum amplitude for a phase slip $\upsilon$ introduced
above, i.e. we assume that $I\gg I_1=I^0_c\upsilon/\pi E_J$.
In this case transitions within each rhombus between the states corresponding
to different minima of the potential~(\ref{fission_potential}) are incoherent.
Total voltage along the chain can be expressed
in terms of $\tau$ as $V = N
\hbar\,\overline{\dot{\theta}}/2e\approx\pi N\hbar/2e\tau$ since
during each jump of the system from one minimum to another the
phase $\theta$ changes by~$\pi$. Thus we obtain for low-current
$V(I)$ dependence:
\begin{equation}
V(I)=\frac{\pi NE_J}{e}\left(\frac{E_C}{E_J}\right)^{1/4}
K(I/I_c^0)\exp\left(-2\sqrt{\frac{E_J}{E_C}}\,s(I/I_c^0)\right)
\label{V(I)}
\end{equation}

\begin{figure}
\includegraphics[width=300pt]{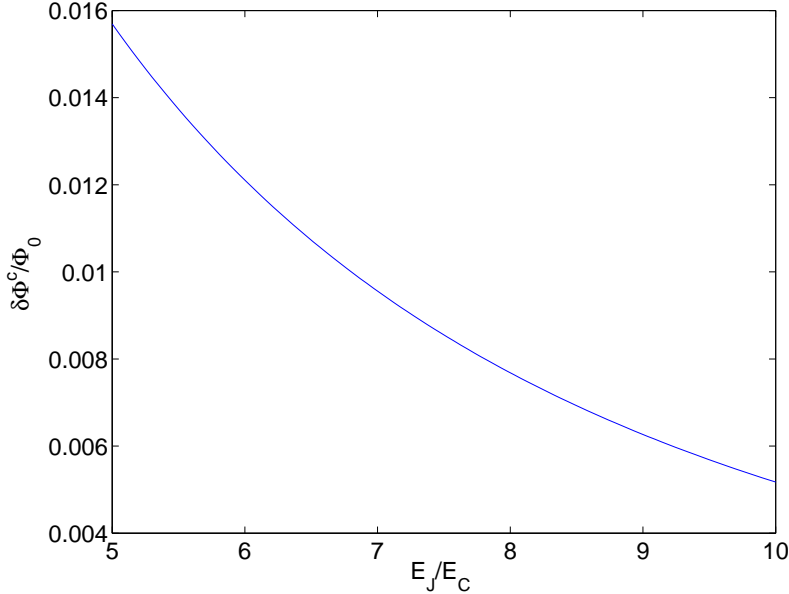}
\caption{\small The critical deviation $\delta\Phi^c$ as a
function of ratio $E_J/E_C$} \label{delta_Phi}
\end{figure}

Equation~(\ref{V(I)}) describes the rhombi chain when the bias current $I$ is
large enough: $I>I_c$, $I\gg I_1$. Under these conditions  coherence in the
system is  destroyed. This  limit is opposite to the one we have considered
in previous Section, where the value of equilibrium Josephson current was determined
by  {\it coherent} quantum fluctuations of all rhombi.

\section{Conclusions}

In this paper we provide detailed calculations of superconductive
current in a long chain composed of frustrated rhombi (i.e. loops
made of 4 superconductive islands).  We show that supercurrent
carried in $4e$ quanta dominates over usual $2e$ supercurrent in
the close vicinity of the maximally frustrated point
$\Phi_r=\Phi_0/2$. According to~(\ref{crossover}) the critical
deviation $\delta\Phi^c$ from this point, which brings the system
back to usual $2e$-supercurrent, depends on the only
parameter~$E_J/E_C$. This dependence is presented on
Fig.~\ref{delta_Phi}. We see that $\delta\Phi^c$ rapidly decreases
with the increase of the ratio $E_J/E_C$. In order to observe
experimentally the $4e$-supprecurent one should control the flux
$\Phi_r$ penetrating each rhombus with accuracy better than
$\delta\Phi^c$. Thus $E_J/E_C$ should not be too large.
\begin{figure}
\includegraphics[width=300pt]{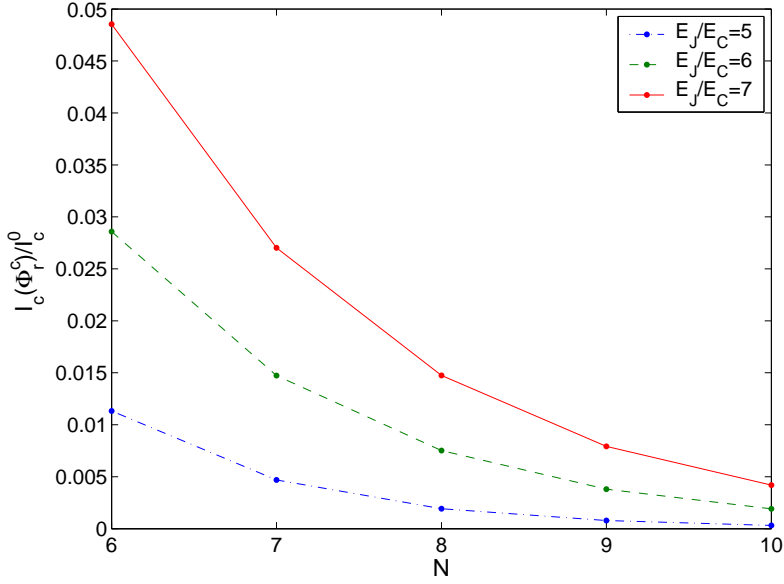}
\caption{\small The critical current $I_c$ at the crossover point as a function of $N$
at differnt $E_J/E_C$.}
\label{I_c_cross}
\end{figure}

In accordance with~(\ref{bwh}) the parameter $q=Nw/2$ governing
the strength of fluctuations in the maximally frustrated point is
proportional to $N^2\upsilon/E_J$ with $\upsilon$ defined by
equation~(\ref{upsilon_gen}). In the regime of strong fluctuations
(large $q$) both $4e$- and $2e$-supercurrents are exponentially
small, cf. Eq.~(\ref{components}). The actions $S_E^{4e}$ and
$S_E^{2e}$ in~(\ref{components}) are proportional to the number of
rhombi~$N$: $S_E^{2e}\sim N\delta \widetilde{S}_E^{2e}$ and
$S_E^{4e}\sim N\delta \widetilde{S}_E^{4e}$. At sufficiently large
$N$ small variations of $\delta\widetilde{S}_E^{4e}$ and
$\delta\widetilde{S}_E^{2e}$ near the crossover point $\Phi_r^c$
lead to strong alteration in the relative weight of $2e$- and
$4e$-supercurrents. Thus  the crossover between $2e$- and
$4e$-regimes is expected to be sharp for large $N$ and $N^2
\upsilon/E_J \geq 1$. On the other hand, the magnitudes of
supercurrent components $I_{4e}$ and $I_{2e}$, although suppressed
by quantum fluctuations, should be not too weak to be measured.
The semi-qualitative dependence of the critical current $I_c$ at
the crossover point on the number of rhombi at different $E_J/E_C$
is presented on Fig.~(\ref{I_c_cross}). While calculating the
curves depicted on Fig.~\ref{I_c_cross} the pre-exponential factor
in the expression for the critical current was evaluated as a
geometrical mean of the prefactors in~(\ref{I2e}) and~(\ref{I4e}).
The optimal set of parameters seems to be the following: $5 \leq
E_J/E_C \leq 7 $, and $ N \in (6,10) $. According to
Figs.~\ref{delta_Phi} and \ref{I_c_cross} it would give
$\delta\Phi^c/\Phi_0 \in (0.01-0.015) $ and $I_c \sim 10^{-2}
-10^{-3} I_{c}^0$.

In this paper we have analysed the path
integral~(\ref{basic_path_integral_for_spin_system_for_lin}) for the limit of relatively
large $\delta$ when $h\gg w$. For this condition to hold at the crossover point we
need, cf.~(\ref{bwh}, \ref{crossover})
\begin{equation}
\left.\frac{h}{w}\right|_{\Phi_r=\Phi_r^c}=
\left(\frac{\pi^2}{64\sqrt{2}d_0^2}\frac{E_J}{\upsilon}\right)^{1/3}\approx
0.2\left(\frac{E_J}{\upsilon}\right)^{1/3}\gg 1
\end{equation}
This is always true for large $E_J/E_C$. However for the proposed
set of parameters ($5\leq E_J/E_C\leq 7$) the ratio $0.75\leq
h/w\leq 0.95$ and we are at the edge of the validity region for
our approximation. Therefore in order to obtain accurate estimates
for $I_{2e}$ and $I_{4e}$ at the point of crossover one needs to
calculate the classical actions on $2e$- and $4e$-trajectories for
the full path
integral~(\ref{basic_path_integral_for_spin_system_for_lin}).
\begin{figure}[t] \centering
\includegraphics[width=300pt]{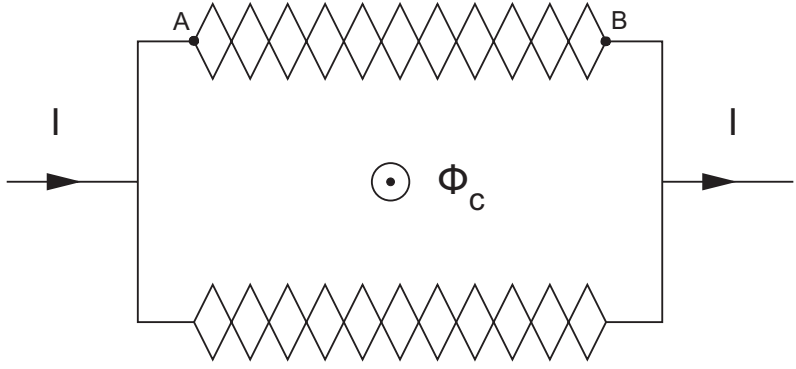}
 \caption{\small An analog of dc-SQUID configuration for the measurement of the current-phase relation~(\ref{gen_cur}).}
\label{exprelease}
\end{figure}

\begin{figure}[t] \centering
\includegraphics[width=300pt]{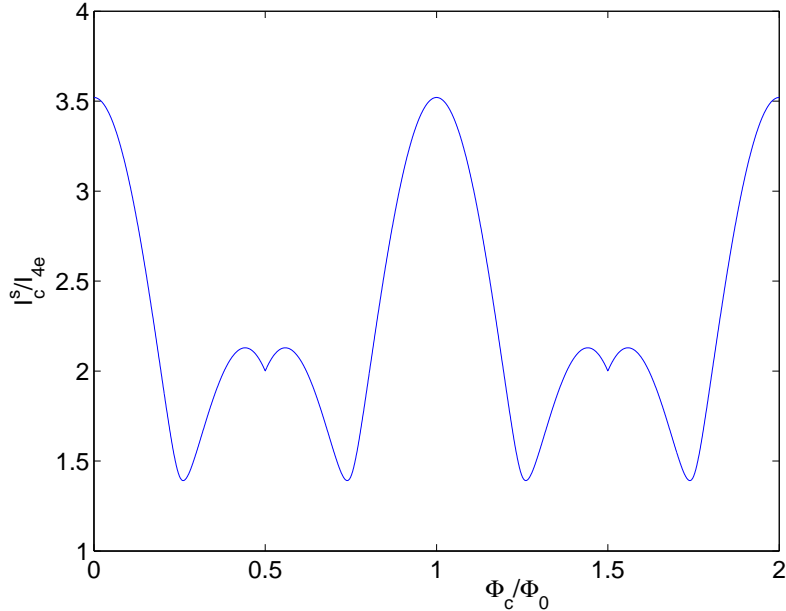}
 \caption{\small The critical current $I_c^s$ at the crossover point.}
\label{I_c^s}
\end{figure}

Possible experimental arrangement for testing the current-phase
relation~(\ref{gen_cur}) is presented on Fig.~\ref{exprelease}.
The circuit on Fig.~\ref{exprelease} is an analog of a simple
dc-SQUID. Let us denote by $\phi$ the order parameter phase
difference in points $A$ and $B$. It follows from
Eq.~(\ref{gen_cur}), that in order to evaluate the critical
current of the proposed device, much as with a dc-SQUID, one needs
to maximize over phase $\phi$ the current $I$ given by
\begin{equation}
I=2I_{4e}\cos\frac{2\pi\Phi_c}{\Phi_0}\sin\left(2\phi+\frac{2\pi\Phi_c}{\Phi_0}\right)+
2I_{2e}\cos\frac{\pi\Phi_c}{\Phi_0}\sin\left(\phi+\frac{\pi\Phi_c}{\Phi_0}\right)\,.
\end{equation}
When the deviation $\delta\Phi$ of the magnetic flux $\Phi_r$
through each rhombus from $\Phi_0/2$ exceeds the critical
deviation $\delta\Phi^c$, the $4e$-supercurrent is negligible and
for the critical current of the circuit on Fig.~(\ref{exprelease})
we get (in complete analogy with a dc-SQUID)
\begin{equation}
I_c^s=2I_{2e}\left|\cos\frac{\pi\Phi_c}{\Phi_0}\right|\,.
\label{I_c^s_large_delta}
\end{equation}
So the dependence of $I_c^s$ on the flux $\Phi^c$ is
$\Phi_0$-periodic for $\delta\Phi\gg\delta\Phi_c$. On the other
hand, at the maximally frustrated point only $4e$-supercurrent
survives so that $I_c^s$ is $\Phi_0/2$-periodic
\begin{equation}
I_c^s=2I_{4e}\left|\cos\frac{2\pi\Phi_c}{\Phi_0}\right|\,.
\label{I_c^s_max_frust}
\end{equation}
The dependence $I_c^s(\Phi_c)$ at the crossover point
($I_{2e}=I_{4e}$) is presented on Fig.~\ref{I_c^s}.

In our analysis we neglected two intrinsic sources of disorder
which are always present in the problem considered: a) some weak
randomness of fluxes $\Phi_r^j$ penetrating different rhombi (due
to unavoidable differences in their areas), and b) random stray
charges $q_n$ which produce, due to Aharonov-Casher effect, some
random phase factors  to the phase slip tunnelling amplitudes.
Whereas the effect of type-a) disorder may be expected to be weak
if areas of different rhombi coincide with the accuracy better
than $\delta\Phi^c/\Phi_0$, the b)-type effect  may occur to be
more severe, cf. Ref.~\cite{Larkin}, where it was discussed for
the simple JJ chain.  We plan to study these effects in further
publications.

We are grateful to L.~B.~Ioffe and B.~Pannetier for many useful and inspiring
discussions and sharing with us their unpublished results.
This research was supported by  the Program ``Quantum Macrophysics" of
the Russian Academy of Sciences, Russian Ministry of Science and  RFBR under
grants No.\ 04-02-16348.  I.V.P. acknowledges financial support from the Dynasty Foundation.

\appendix
\section{Semiclassical analysis for $I(\gamma)$}

In this Appendix we will obtain the dependence of the lowest eigenvalue $b$ of the
problem defined by Eqs.(\ref{boundary},\ref{main_res_under_flux_dif_from_pi} ) on the
phase $\widetilde{\gamma}$ in the regime of strong fluctuations and derive the
expression~(\ref{gen_cur}) for the persistent current.

\begin{figure}
\includegraphics[width=300pt]{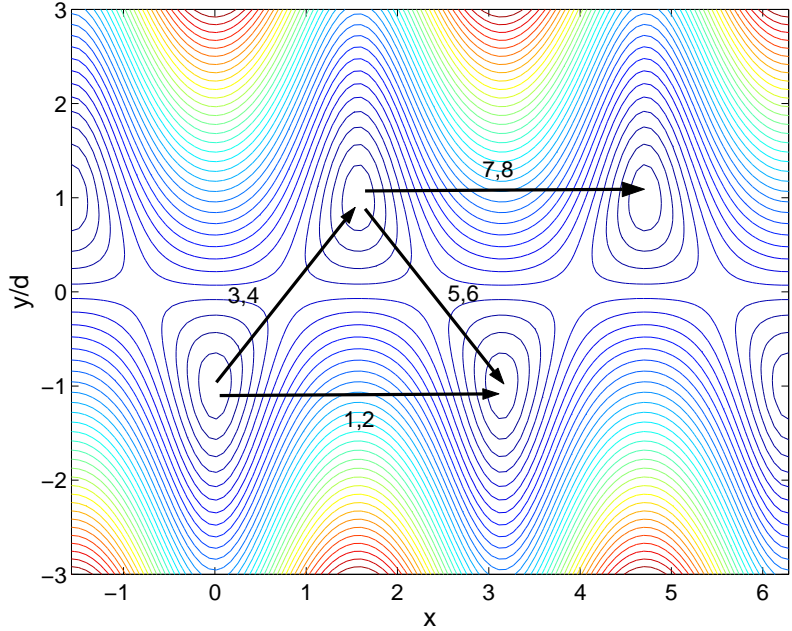}
\caption{\small The contour plot of the potential $U_{eff}(x,y)$.}
\label{U_eff}
\end{figure}

Let us analyse the transition amplitude~(\ref{2D_path_integral})
in more details for the case when $(x_1,y_1)$ is an "even" minimum
of the potential~(\ref{effektive_potential}) and $(x_2, y_2)$ is
the neares "odd" one, cf.~(\ref{even},\ref{odd}). To be specific
we choose $(x_1,y_1)=(0, -d)$ and $(x_2,y_2)=(\pi/2, d)$. The
contour plot of the potential $U_{eff}(x,y)$ is presented on
Fig.~\ref{U_eff}. Possible tunneling trajectories of the system
are schematically depicted with arrows. It is convenient to divide
all trajectories into eight groups. Along each trajectory from
group $1$, variable  $y$ is unchanged and equals $-d$ on both ends
of the trajectory, whereas variable $x$ inreases by $\pi$; a
trajectory from group $2$ is a counterpart (going  against
the arrow on Fig.~\ref{U_eff}) to the previous one.
The groups $3,\,4,\dots,8$ are defined in the
same way according to Fig.~\ref{U_eff}.
All trajectories from groups
$1,2,7,8$ connects minima of the same parity and so are of the first type according
to Section $3$, whereas the trajectories from groups
$3,4,5,6$ connects minima of opposite parity and are of the second type.

Let us denote by $T\alpha_{4e}$ and $T\alpha_{2e}$  contributions to the tunneling
amplitude from a {\it single} trajectory of the first and the second type respectively,
i.e.
\begin{equation}
\alpha_{4e}=\beta_{4e}e^{-S_E^{4e}} \qquad {\rm and}\ \qquad
\alpha_{2e}=\beta_{2e}e^{-S_E^{2e}}\,,
\end{equation}
where the prefactors $\beta_{4e}$ and $\beta_{2e}$ can be obtained by integration over
the fluctuations near the corresponding trajectories. In order to evaluate the transition
amplitude~(\ref{2D_path_integral}) in semiclassical approximation one should sum up the
contributions from all trajectories consisting of $n_1$ subtrajectories from group 1,
$n_2$ subtrajectories from group 2 and so on. Such a trajectory including
$R=\sum_{k=1}^{8}n_k$ subtrajectories gives to the path integral~(\ref{2D_path_integral})
\begin{equation}
\frac{T^R}{R!}\alpha_{4e}^{n_3+n_4+n_5+n_6}\alpha_{2e}^{n_1+n_2+n_7+n_8}
\label{contr}
\end{equation}

As the trajectory under consideration starts at $(0, -d)$ and ends at $(\pi/2, d)$
one should impose  two additional constraints upon the integers  $n_1,\dots,n_8$ :
\begin{equation}
2(n_1+n_7)-2(n_2+n_8)+n_3-n_4+n_5-n_6=1\,,
\label{cond_1}
\end{equation}
\begin{equation}
n_3-n_4-n_5+n_6=1\,.
\label{cond_2}
\end{equation}
Let us introduce
\begin{equation}
K=n_1+n_7\,, \qquad L=n_2+n_8\,\qquad {\rm and}\qquad M=n_4+n_5=n_3+n_6-1\,.
\end{equation}
All trajectories with fixed $K$, $L$, $M$, $n_3$ and $n_4$ give the same contribution to
the transition amplitude. The number of such trajectories is
\begin{equation}
\frac{R!(M+1)!M!}{(2M+1)!K!L!n_3!n_4!(M-n_4)!(M-n_3+1)!}
\label{number}
\end{equation}
Thus, taking into account~(\ref{contr},\ref{number},\ref{cond_1}),
for the transition amplitude in semiclassical approximation we get
\begin{equation}
\left<y_2\,,x_2\right|e^{-T\widehat{H}}\left|y_1\,,x_1\right>=
\sum_{\substack{K,L,n_3,n_4\geq0 \\
M\geq max(n_4, n_3-1)\\
K-L+n_3-n_4=1}}
\frac{(T\alpha_{4e})^{K+L}(T\alpha_{2e})^{2M+1}(M+1)!M!}{K!L!n_3!n_4!(2M+1)!
(M-n_4)!(M-n_3+1)!}
\label{main_sum}
\end{equation}
Instead of calculating the sum in~(\ref{main_sum}) it is convenient to evaluate function
$Q(\alpha_{2e}, \alpha_{4e})$ defined by
\begin{equation}
\left<y_2\,,x_2\right|e^{-T\widehat{H}}\left|y_1\,,x_1\right>=
\frac{1}{T}\frac{\partial Q}{\partial \alpha_{2e}}
\label{def_Q}
\end{equation}
For the function $Q(\alpha_{2e}, \alpha_{4e})$ we have
\begin{equation}
Q(\alpha_{2e}, \alpha_{4e})=\frac{1}{2} \sum_{\substack{K,L,n_3,n_4\geq0 \\
M\geq max(n_4, n_3-1)\\
K-L+n_3-n_4=1}}
\frac{(T\alpha_{4e})^{K+L}(T\alpha_{2e})^{2M+2}B(M+1,M+1)}{K!L!n_3!n_4!
(M-n_4)!(M-n_3+1)!}
\label{Q_sum}
\end{equation}
where $B(x,y)$ is Euler's beta function. Using the integral representation for the
beta function and the ascending series for the modified Bessel
function~\cite{Abramowitz}
\begin{equation}
B(x,y)=2\int_0^{\pi/2}d\varphi(\sin\varphi)^{2x-1}(\cos\varphi)^{2y-1}\,,
\qquad
I_n(z)=\left(\frac{1}{2}z\right)^n\sum_k \frac{\left(\frac{1}{4}z^2\right)^k}{k!(n+k)!}
\label{BI}
\end{equation}
we can carry out the summation over $M$, $K$, $L$ in~(\ref{Q_sum}) and obtain
\begin{equation}
Q=\sum_{n_3, n_4\geq0}\frac{(T\alpha_{2e})^{n_4+n_3+1} I_{|n_4-n_3+1|}(2T\alpha_{4e})}
{2^{n_3+n_4+1}n_3!n_4!}\int_0^{\pi}d\varphi (\sin\varphi)^{n_3+n_4}
I_{|n_4-n_3+1|}(T\alpha_{2e}\sin\varphi)\,.
\end{equation}
Introducing $Z=n_3-1-n_4$ and accomplishing the summation over $n_3$, $n_4$ under
fixed $Z$ we get
\begin{equation}
Q=\frac{T\alpha_{2e}}{2}\sum_{Z=-\infty}^{+\infty}\int_0^{\pi}d\varphi
I_{|Z|}(2T\alpha_{4e})I_{|Z|}(T\alpha_{2e}\sin\varphi)
I_{|Z+1|}(T\alpha_{2e}\sin\varphi)\,.
\label{sum_over_Z}
\end{equation}
Taking into account the integral representation of the modified Bessel function
and its generating function~\cite{Abramowitz}
\begin{equation}
I_n(z)=\frac{1}{2\pi}\int_0^{2\pi}d\theta \exp\left(in\theta+z\cos\theta\right)\,,
\qquad
\exp\left\{\frac{z}{2}\left(t+\frac{1}{t}\right)\right\}=
\sum_{k=-\infty}^{+\infty}t^kI_k(z)
\end{equation}
we can rewrite~(\ref{sum_over_Z}) in the form
\begin{multline}
Q(\alpha_{2e},\alpha_{4e})=\frac{T\alpha_{2e}}{4}
\int_0^{\pi} d\varphi \int_0^{2\pi}\frac{d\theta_1 d\theta_2}{(2\pi)^2}
(\cos\theta_1+\cos\theta_2)\times\\
\exp\left\{T\alpha_{2e}\sin\varphi(\cos\theta_1+\cos\theta_2)+
2T\alpha_{4e}\cos(\theta_1+\theta_2)\right\}\,.
\end{multline}
After the substitution $u=(\theta_1+\theta_2)/2$, $v=(\theta_1-\theta_2)/2$ we integrate
over $\varphi$ and $v$ using the relation~\cite{Gradstein}
\begin{equation}
\int_0^{\pi/2}d\varphi I_1(2z\sin\varphi)=\frac{\pi}{2}I_{1/2}^2(z)=\frac{\sinh^2 z}{z}
\end{equation}
and derive an integral representation for $Q(\alpha_{2e},\alpha_{4e})$
\begin{equation}
Q(\alpha_{2e},\alpha_{4e})=\int_0^{\pi}\frac{du}{2\pi}\left\{
\cosh\left(2T\alpha_{2e}\cos u\right)-1\right\}
\exp\left(2T\alpha_{4e}\cos2u\right)\,.
\end{equation}
Finally, with the aid of~(\ref{def_Q}) we obtain an explicit expression for the
semiclassical transition amplitude~(\ref{2D_path_integral})
\begin{equation}
\left<y_2\,,x_2\right|e^{-T\widehat{H}}\left|y_1\,,x_1\right>=
\int_0^{2\pi}\frac{du}{2\pi}e^{iu}\exp(2T\alpha_{4e}\cos 2u+2T\alpha_{2e}\cos u)
\label{xy_tr_amp_final}
\end{equation}

On the other hand the transition amplitude~(\ref{2D_path_integral}) can be written as
a sum over the system's eigenstates
\begin{equation}
\left<y_2\,,x_2\right|e^{-T\widehat{H}}\left|y_1\,,x_1\right>=
\sum_n\psi_n^*(x_1,y_1)\psi_n(x_2,y_2)e^{-E_nT}\,.
\label{xy_tr_amp_expan}
\end{equation}
Note that the symmetry group of the potential $U_{eff}(x,y)$ consist of transformations
$\widehat{V}_n=\widehat{R}^n\widehat{T}_{\pi n/2}$, where $\widehat{T}_a$ is operator
of the translation over distance $a$ along the $x$-axis introduced in Section~$3$ and
$\widehat{R}$ is operator of the reflection in the $x$-axis.

Thus the energy levels $E_u^0$ of a fictitios particle moving in the potential
$U_{eff}$ are classified by imposing on their wave functions $\psi_u$ a twisted
boundary condition
\begin{equation}
\widehat{V_1}\psi_u(x,y)\equiv\psi_u(x+\pi/2, -y)=e^{iu}\psi_u(x,y)\,.
\label{xy_boundary}
\end{equation}
Comparing~(\ref{xy_tr_amp_expan}, \ref{xy_boundary}) with~(\ref{xy_tr_amp_final})
we conclude that the result~(\ref{xy_tr_amp_final}) has the form of the
expansion~(\ref{xy_tr_amp_expan}) with the multiple $e^{iu}$ under the integral
emerging from $\psi^*_u(x_1,y_1)\psi_u(x_2,y_2)$ and the remaining part of the
under-integral expression providing us with the particle energy
\begin{equation}
E_u^0= -2\alpha_{4e}\cos2u-2\alpha_{2e}\cos u\,.
\end{equation}

Coming back to the original problem defined by Eqs.~(\ref{boundary},
\ref{main_res_under_flux_dif_from_pi}) and comparing~(\ref{boundary})
with~(\ref{xy_boundary}) we see that we should identify the phase
$\widetilde{\gamma}$ with the "quasimomentum" $u$.
Taking into account the relation between $b$ and the energy of the fictitios
particle $b=2E^0$ mentioned in Section 3 we finally obtain the
$b(\widetilde{\gamma})$ dependence:
\begin{equation}
b(\widetilde{\gamma})=-4\alpha_{4e}\cos2\widetilde{\gamma}-
4\alpha_{2e}\cos\widetilde{\gamma}\,.
\label{gen_b(gamma)}
\end{equation}
With  Eq.~(\ref{gen_b(gamma)}) and standard relation
$I(\gamma) = (2e/\hbar) dE_0/d\gamma$
we easily recover the results~(\ref{gen_cur}) and~(\ref{components}).


\begin{thebibliography}{8}
\bibitem{Doucot} B. Doucot and J. Vidal, Phys. Rev. Lett. {\bf 88}, 227005
(2002).
\bibitem{IoffeFeigelman02} L. B. Ioffe and M. V. Feigel'man, Phys. Rev. B {\bf 66}, 224503 (2002)
\bibitem{Doucot03} B. Doucot, M. V. Feigel'man and L. B. Ioffe, Phys. Rev. Lett.{\bf 90}, 107003  (2003)
\bibitem{capacitances} P. Delsing, C. D. Chen, D. B. Halivald, Y. Harada, T. Claeson,
Phys. Rev. B {\bf 50}, 3959 (1994)
\bibitem{Larkin} K. A. Matveev, A. I. Larkin and L. I. Glazman,
Phys. Rev. Lett. {\bf 89}, 096802 (2002).
\bibitem{Ioffe_sp} L. B. Ioffe, private communication.
\bibitem{Abramowitz} M. Abramowitz and I. A. Stegun, {\it Handbook of Mathematical
Functions}, (Dover, New York, 1974).
\bibitem{Klauder}  J. R. Klauder, Phys. Rev. D {\bf 19}, 2349 (1979)
\bibitem{Pannetier} B. Pannetier, private communication
\bibitem{Coleman}  C. Callan, S. Coleman, Phys. Rev. D {\bf 16}, 1762 (1977)
\bibitem{Larkin_Ovchinnikov} A. I. Larkin, U. N. Ovchinnikov, Sov. Phys. JETP {\bf 59},420 (1984)
\bibitem{Gradstein} I. Gradstein, I. Ryzhik, {\it Table of Integrals, Series, and
Products}, (Academic Press, New York, 1980).
\end{thebibliography}
\end{document}